\documentclass[twocolumn]{aastex631}  

\usepackage[utf8]{inputenc} 
\usepackage[T1]{fontenc}    
\usepackage{hyperref}       
\usepackage{url}            
\usepackage{booktabs}       
\usepackage{amsfonts}       
\usepackage{nicefrac}       
\usepackage{microtype}      
\usepackage{lipsum}
\usepackage{amsmath}
\usepackage{amssymb}
\usepackage{multirow}
\usepackage{float}
\usepackage{hyperref}
\usepackage{graphicx}
\usepackage[noabbrev]{cleveref}
\usepackage{physics}

\def\apjl{ApJ}
\def\apjs{ApJS}
\def\ao{Appl.~Opt.}
\def\aap{A\&A}

\newcommand{\tilxi}{\Tilde{\xi}}

\newcommand{\be}{\begin{equation}}
\newcommand{\ee}{\end{equation}}
\newcommand{\bary}{\begin{eqnarray}}
\newcommand{\eary}{\end{eqnarray}}

\begin{document}

\title[Polarization]{Afterglow Polarization from Off-Axis GRB Jets}

\author[0000-0001-5785-8305 ]{A. C. Caligula Do E. S. Pedreira}
\affiliation{Instituto de Astronom\'ia, Universidad Nacional Aut\'onoma de M\'exico, Circuito Exterior, C.U., A. Postal 70-264, 04510, CDMX, Mexico}
\author[0000-0002-0173-6453]{N. Fraija}
\affiliation{Instituto de Astronom\'ia, Universidad Nacional Aut\'onoma de M\'exico, Circuito Exterior, C.U., A. Postal 70-264, 04510, CDMX, Mexico}
\author[0000-0001-5193-3693]{A. Galvan-Gamez} 
\affiliation{Instituto de Astronom\'ia, Universidad Nacional Aut\'onoma de M\'exico, Circuito Exterior, C.U., A. Postal 70-264, 04510, CDMX, Mexico}
\author[0000-0002-2516-5739]{B. Betancourt Kamenetskaia}
\affiliation{TUM Physics Department, Technical University of Munich, James-Franck-Str, 85748 Garching, Germany}
\author[0000-0002-2149-9846]{P. Veres}
\affiliation{Center for Space Plasma and Aeronomic Research (CSPAR), University of Alabama in Huntsville, Huntsville, AL 35899, USA}
\author[0000-0003-4442-8546]{M.G. Dainotti}
\affiliation{Division of Science, National Astronomical Observatory of Japan, 2-21-1 Osawa, Mitaka, Tokyo 181-8588, Japan}
\affiliation{The Graduate University for Advanced Studies (SOKENDAI),
2-21-1 Osawa, Mitaka, Tokyo 181-8588, Japan}
\affiliation{Space Science Institute, 4750 Walnut Street, Boulder, CO
80301, USA}
\affiliation{SLAC National Accelerator Laboratory, 2575 Sand Hill Road, Menlo Park, CA 94025, USA}
\author[0000-0001-6849-1270]{S. Dichiara}
\affiliation{Department of Astronomy, University of Maryland, College Park, MD 20742-4111, USA}
\affiliation{Astrophysics Science Division, NASA Goddard Space Flight Center, 8800 Greenbelt Rd, Greenbelt, MD 20771, USA}
\author[0000-0002-0216-3415]{R.~L.~Becerra}
\affiliation{Instituto de Ciencias Nucleares, Universidad Nacional Aut\'onoma de M\'exico, Apartado Postal 70-264, 04510 M\'exico, CDMX, Mexico}

\shortauthors{A. Pedreira et al.}

\begin{abstract}
{As we further our studies on Gamma-ray bursts (GRBs), both on theoretical models and observational tools, more and more options begin to open for exploration of its physical properties. As transient events primarily dominated by synchrotron radiation, it is expected that the synchrotron photons emitted by GRBs should present some degree of polarization throughout the evolution of the burst. Whereas observing this polarization can still be challenging due to the constraints on observational tools, especially for short GRBs, it is paramount that the groundwork is laid for the day we have abundant data. In this work, we present a polarization model linked with an off-axis spreading top-hat jet synchrotron scenario in a stratified environment with a density profile $n(r)\propto r^ {-k}$. We present this model's expected temporal polarization evolution for a realistic set of afterglow parameters constrained within the values observed in the GRB literature for four degrees of stratification $k=0,1,1.5 {\rm \, and\,} 2$ and two magnetic field configurations with high extreme anisotropy. We apply this model and predict polarization from a set of GRBs exhibiting off-axis afterglow emission. In particular, for GRB 170817A, we use the available polarimetric upper limits to rule out the possibility of a extremely anisotropic configuration for the magnetic field.}
\end{abstract}

\keywords{polarization; grbs; synchrotron; particle acceleration; magnetic fields;}


\section{Introduction}\label{sec1}

Gamma-ray bursts (GRBs) are the most luminous phenomena in the Universe. They originate from the deaths of massive stars \citep{1993ApJ...405..273W,1998ApJ...494L..45P, 2006ARA&A..44..507W,Cano2017} or the merging of two compact objects, like neutron stars \citep[NSs;][]{1989Natur.340..126E, 1992ApJ...392L...9D, 1992Natur.357..472U, 1994MNRAS.270..480T, 2011MNRAS.413.2031M} or a NS with a black hole \citep[BH,][]{1992ApJ...395L..83N}. GRBs are analyzed according to their phenomenology observed during the early and late phases and generally described through the fireball model \citep{1998ApJ...497L..17S} to differentiate the distinct origins. The early and main emission, called the ``prompt emission". is observed from hard X-rays to $\gamma$-rays and explained through interactions of internal shells of material thrown violently from the central engine at different velocities. The late emission, called  ``afterglow" \citep[e.g.,][]{1997Natur.387..783C, 1998ApJ...497L..17S,2002ApJ...568..820G, 1997Natur.386..686V,1998A&A...331L..41P,Gehrels2009ARA&A,Wang2015}, corresponds to the long-lasting multi-wavelength emission observed in gamma-rays, X-rays, optical, and radio. 
The afterglow is usually modelled with synchrotron emission generated when the relativistic outflow transfers a significant fraction of its energy to the external medium.
 GRBs are usually classified as long GRBs (lGRBs) and short (sGRBs), depending on their duration: $T_{90}\leq 2\mathrm{\,s}$ or $T_{90} \ge 2\mathrm{\,s}$,\footnote{$T_{90}$ is the time over which a GRB releases from $5\%$ to $95\%$ of the total measured counts.} respectively \citep{mazets1981catalog, kouveliotou1993identification}.

It is thought that the primary emission mechanism in GRB afterglows is synchrotron emission \citep{Kumar,1997ApJ...476..232M}.  This synchrotron emission, arising from radiating electrons at the forward shock, is dependent on the local magnetic field. The magnetic field behind the shock can originate from the compression of an existing magnetic field within the interstellar medium \citep{1980MNRAS.193..439L,2021MNRAS.507.5340T} and from shock-generated two-stream instabilities \citep{PhysRevLett.2.83, Medvedev}. The interstellar medium magnetic field can be composed of multiple components: a large scale coherent component, a small scale random component, and a striated component that changes directions randomly on small scales but remains aligned over large scales \citep{2018JCAP...08..049B}; while the magnetic field generated by plasma instabilities is random in orientation but mostly confined to the plane of the shock \citep{2020MNRAS.491.5815G}.  There is a tremendous challenge in pinning down the source and configuration of those fields and other physical parameters of GRBs through modeling. This has led to the necessity of other avenues of exploration of these complex systems. One such means is linear polarization.

Synchrotron radiation is naturally polarized. The flux of synchrotron photons emitted throughout the shock peaks on gamma-rays in seconds, on lower frequencies in minutes to hours (e.g., optical bands), eventually reaching radio after a day. Linear polarization has been measured, up to a few percent, from the afterglow of several GRBs. 
Some examples are GRB 191221B ($\Pi=1.2\%$, \cite{Buckley}) for the late afterglow, GRB 190114C ($\Pi=0.8\pm0.13\%$, \cite{Laskar_2019}) on the radio band, and the upper limits determinations of GRB 991216  \citep[yielding $\Pi < 7\%$, ][]{2005ApJ...625..263G} and GRB 170817A \citep[yielding $\Pi < 12\%$, on the 2.8GHz radio band][]{2018ApJ...861L..10C}. 
Since the degree of polarization is intrinsically dependent upon the configuration of the magnetic field and jet structure, analysis of the polarization degree across all epochs of the GRB allows us to look further into these configurations and, consequently, their sources.   Many researchers, such as \cite{2003ApJ...594L..83G, Gill-1, 2004MNRAS.354...86R, Lyutikov, Nakar, 2021MNRAS.507.5340T, 2020ApJ...892..131S}, have already addressed their investigation on the viability of using polarization models to obtain information related to the source. One of the most significant obstacles has been the scarcity of polarization data for GRBs due to the unfortunate small number of orbital polarimeters and the typical difficulties in observing these luminous events. Despite that, advances have been made in the area, and thanks to efforts like the POLAR project \citep{POLAR}, it is expected that in future years we should have an abundance of data for the test of different models.

This work extends the analytical synchrotron afterglow scenario, of the off-axis jet in a stratified environment used to describe the multi-wavelength observations in GRB 170817A, and a sample of some GRBs showing off-axis emission with similar characteristics. We present, in general, the temporal evolution of polarization from the synchrotron afterglow stratified model and compute the expected polarization for bursts previously modeled by an off-axis emission: GRB 080503 \citep{2009ApJ...696.1871P,2015ApJ...807..163G}, GRB 140903A \citep{2016ApJ...827..102T, 2017ApJ...835...73Z}, GRB 150101B \citep{2018NatCo...9.4089T}, GRB 160821B \citep{2019MNRAS.489.2104T}, GRB 170817A \citep{2017Sci...358.1559K, 2017MNRAS.472.4953L, mooley, 2018ApJ...867...95H, 2019ApJ...884...71F} and SN2020bvc \citep[also, see][for a more detailed discussion on the modeling of these events]{2022arXiv220502459F} -- which is thought to be linked to an off-axis GRB \citep{2020A&A...639L..11I}. In particular, for GRB 170817A, we use the available polarimetric upper limits.  With this in mind, the structure of the paper is as follows. In Section \ref{sec2}, we briefly show the off-axis jet synchrotron model derived in  \cite{2022arXiv220502459F}.  In Section \ref{sec3}, we introduce the polarization model to be utilized throughout this paper. In Section \ref{sec4}, we compute the expected polarization and present the results for a sample of bursts showing off-axis afterglow emission. Finally, in Section \ref{sec7}, we summarize our work and offer our concluding remarks. 

\section{Synchrotron Polarization from an off-axis top-hat Jet}\label{sec2}

In the following section, we present the off-axis equations of the synchrotron scenario presented in \cite{2022arXiv220502459F}, which is applied to the polarization model for time-evolving calculations.

\subsection{Synchrotron scenario}\label{subsec21}

In forward-shock models, accelerated electrons are described by taking into account their Lorentz factors ($\gamma_e$) and the electron power index $p$. This leads to a distribution of the form $N(\gamma_e)\,d\gamma_e \propto \gamma_e^{-p}\,d\gamma_e$ for $\gamma_m\leq \gamma_{\rm e}$, where $\gamma_m=m_{\rm p}/m_{\rm e}g(p)\varepsilon_{\rm e}(\Gamma-1)\zeta^{-1}_e$ is the minimum electron Lorentz factor with $\Gamma$ the bulk Lorentz factor, $m_{\rm p}$ and $m_{\rm e}$ the proton and electron mass, respectively, $\varepsilon_{\rm e}$ the fraction of energy given to accelerate electrons, $\zeta_{e}$ the fraction of electrons that were accelerated by the shock front \citep{2006MNRAS.369..197F} and $g(p)=\frac{p-2}{p-1}$. The comoving magnetic field strength in the blast wave can be expressed as  $B'^2/(8\pi)=\varepsilon_Be$, where knowledge of the energy density  $e=[(\hat\gamma\Gamma +1)/(\hat\gamma - 1)](\Gamma -1)n(r) m_pc^2$, adiabatic index $\hat\gamma$ \citep{1999MNRAS.309..513H} and fraction of energy provided to the magnetic field ($\varepsilon_B$) is necessary. In what follows, we adopt the unprimed and prime terms to refer them in the observer and  comoving frames, respectively. 

In this work, we will consider the evolution of the forward shock in a stratified medium. To this end, we model the surrounding number density as $n(r)=A_{\rm k} r^{\rm -k}$ with $A_{\rm k} =n_{\rm 0}(r_0) \, r_0^{\rm k}$, where $n_0$ is the density at initial radius $r_0$. The stratification parameter, $\rm k$, lies in the range $0\leq k < 3$, with ${\rm k=0}$ corresponding to a constant-density medium, and ${\rm k = 2}$ to a stellar wind ejected by its progenitor. The cooling electron Lorentz factor is written as $\gamma_{\rm c}=(6\pi m_e c/\sigma_T)(1+Y)^{-1}\Gamma^{-1}B'^{-2}t^{-1}$, where $\sigma_T$ is the Thomson cross-section and $Y$ is the Compton parameter \citep{2001ApJ...548..787S, 2010ApJ...712.1232W}. The synchrotron spectral breaks can now be expressed in terms of previously defined quantities as $\nu'_{\rm i}=q_e/(2\pi m_ec)\gamma^{2}_{\rm i}B'$, where the sub-index ${\rm i=m}$ and ${\rm c}$ will stand for the characteristic or cooling break, respectively. The constants $q_e$ and $c$ are the elementary charge and the speed of light, respectively. The synchrotron radiation power per electron in the comoving frame is given by $P'_{\nu'_m}\simeq \sqrt{3}q_e^3/(m_ec^2)B'$ \citep[e.g., see][]{1998ApJ...497L..17S, 2015ApJ...804..105F}.  Considering  the total number of emitting electrons $N_e=(\Omega/4\pi)\, n(r) \frac{4\pi}{3-k} r^3$ and also taking into account the transformation laws for  the solid angle ($\Omega= \Omega'/\delta^2_D$), the radiation power ($P_{\nu_m}=\delta_D/(1+z) P'_{\nu'_m}$) and the spectral breaks  ($\nu_{\rm i}=\delta_D/(1+z)\nu'_{\rm i}$), the maximum flux given by synchrotron radiation is

\be
F_{\rm \nu, max}=\frac{(1+z)^2\delta^3_D}{4\pi d_z^2}N_eP'_{\nu'_m}\,,
\ee
 where {\small $d_{\rm z}=(1+z)\frac{c}{H_0}\int^z_0\,\frac{d\tilde{z}}{\sqrt{\Omega_{\rm M}(1+\tilde{z})^3+\Omega_\Lambda}}$}  \citep{1972gcpa.book.....W}  is the luminosity distance, $r=\delta_D/(1+z) \Gamma\beta c t$ is the shock radius, and $\delta_D=\frac{1}{\Gamma(1-\mu\beta)}$ is the Doppler factor with $\mu=\cos \Delta \theta$, $\beta=v/c$, where $v$ is the velocity of the material, and $\Delta \theta=\theta_{\rm obs} - \theta_{\rm j}$ is given by the viewing angle ($\theta_{\rm obs}$) and the half-opening angle of the jet ($\theta_{\rm j}$). For the cosmological constants, we assume a spatially flat universe $\Lambda$CDM model with  $H_0=69.6\,{\rm km\,s^{-1}\,Mpc^{-1}}$, $\Omega_{\rm M}=0.286$ and  $\Omega_\Lambda=0.714$ \citep{2016A&A...594A..13P}. 

We assume an adiabatic evolution of the forward shock with an isotropic equivalent-kinetic energy   $E=\frac{4\pi}{3-k}r^{3-k}m_p c^2 A_{\rm k}  \Gamma^2$ \citep[Blandford-McKee solution;][]{1976PhFl...19.1130B} and a radial distance $r=c\beta t/[(1+z)(1-\beta\mu)]$. Then, the evolution of the bulk Lorentz factor is given by 

\bary
\label{Gamma_dec_off}
\Gamma &=& \left(\frac{3}{4\pi\,m_p c^{5-k}}\right)^{\frac12} \,(1+z)^{-\frac{k-3}{2}}  (1-\beta\cos\Delta\theta)^{-\frac{k-3}{2}}\,A_{\rm k}^{-\frac{1}{2}}\,\cr
&&\hspace{6cm}\times E^{\frac{1}{2}}t^{\frac{k-3}{2}} \,,
\eary

with $\beta=\sqrt{\Gamma^2-1}/\Gamma$. The deceleration time scale $t_{\rm dec}$ can be defined using Eq. \ref{Gamma_dec_off}. 

During the deceleration phase before afterglow emission enters in the observer's field of view, the bulk Lorentz factor is given by Eq. \ref{Gamma_dec_off}.   The minimum and cooling electron Lorentz factors are  given by
{\small
\bary\label{eLor_syn_ism1}\nonumber
\gamma_m&=& \gamma^0_m\,\left(\frac{1+z}{1.025}\right)^{\frac{3-k}{2}}\zeta_{e}^{-1} A_{k}^{-\frac{1}{2}} \varepsilon_{e,-1}\theta_{j,5}^{-1}\Delta\theta_{15}^{3-k}E_{51}^{\frac{1}{2}}t_{7.0}^{\frac{k-3}{2}}\cr
\gamma_c&=&\gamma^0_c\,\left(\frac{1+z}{1.025}\right)^{-\frac{k+1}{2}} A_{k}^{-\frac{1}{2}} (1+Y)^{-1}\varepsilon_{B,-3}^{-1}  \theta_{j,5}\Delta\theta_{15}^{-(k+1)}E_{51}^{-\frac{1}{2}}\cr 
&&\hspace{6.3cm}\times t_{7.0}^{\frac{k+1}{2}}\,,
\eary
}
respectively, which correspond to a comoving magnetic field given by {\small $B'\propto \,\left(\frac{1+z}{1.025}\right)^{\frac{3}{2}}\varepsilon_{B,-3}^{\frac{1}{2}}\theta_{j,5}^{-1}\Delta\theta_{15}^{3}E_{51}^{\frac{1}{2}} t_{7.0}^{-\frac{3}{2}}$}.
%
%
%
The synchrotron spectral breaks and the maximum flux can be written as
{\small
\bary\label{En_br_syn_off}
\nu_{\rm m}&=& \nu^0_{\rm m}\left(\frac{1+z}{1.025}\right)^{\frac{4-k}{2}}\zeta_{e}^{-2} A_{k}^{-\frac{1}{2}} \varepsilon_{e,-1}^2 \varepsilon_{B,-3}^{\frac{1}{2}}\theta_{j,5}^{-2}\Delta\theta_{15}^{4-k}E_{51} t_{7.0}^{\frac{k-6}{2}}\cr
\nu_{\rm c}&=& \nu^0_{\rm c} \left(\frac{1+z}{1.025}\right)^{-\frac{k+4}{2}} A_{k}^{-\frac{1}{2}} (1+Y)^{-2}\varepsilon_{B,-3}^{-\frac{3}{2}}\theta_{j,5}^{2}\Delta\theta_{15}^{-(k+4)}\cr
&&\hspace{5.8cm}\times E_{51}^{-1} t_{7.0}^{\frac{k+2}{2}}\cr 
F_{\rm max} &=& F^0_{\rm max}\left(\frac{1+z}{1.025}\right)^{\frac{5k-8}{2}}\zeta_{e} A_{k}^{\frac{5}{2}} \varepsilon_{B,-3}^{\frac{1}{2}}d_{z,26.5}^{-2}\theta_{j,5}^{2}\Delta\theta_{15}^{5k-18}\cr
&&\hspace{5.2cm}\times E_{51}^{-1} t_{7.0}^{\frac{12-5k}{2}}\,,
\eary
}
respectively.  The synchrotron spectral breaks in the  self-absorption regime are derived from  $\nu'_{\rm a,1}=\nu'_{\rm c}\tau^{\frac35}_{0,m}$,  $\nu'_{\rm a,2}=\nu'_{\rm m}\tau^{\frac{2}{p+4}}_{0,m}$ and $\nu'_{\rm a,3}=\nu'_{\rm m}\tau^{\frac35}_{0,c}$ with the optical depth given by $\tau_{0,i}\simeq\frac{5}{3-k}\frac{q_en(r)r}{B'\gamma^5_{\rm i}}$, with $r$ the shock radius \citep{1998ApJ...501..772P}. Therefore, the spectral breaks in the self-absorption regime are given by

{\small
\bary\label{SelfAbsorptionCuts_off}
\nu_{\rm a,1}&\simeq& \nu^0_{\rm a,1}\left(\frac{1+z}{1.025}\right)^{\frac{4(2k-5)}{5}}\zeta_{e}^{\frac85} A_{k}^{\frac{8}{5}} \varepsilon_{e,-1}^{-1} \varepsilon_{B,-3}^{\frac{1}{5}}\theta_{j,5}^{\frac{8}{5}} \Delta\theta_{15}^{\frac{8(2k-5)}{5}}\cr
&&\hspace{5.8cm}\times E_{51}^{-\frac{4}{5}}t_{7.0}^{\frac{15-8k}{5}}\cr
\nu_{\rm a,2}&\simeq& \nu^0_{\rm a,2} \left(\frac{1+z}{1.025}\right)^{-\frac{24-10k-4p+kp}{2(p+4)}}\zeta_{e}^{\frac{2(2-p)}{p+4}} A_{k}^{\frac{10-p}{2(p+4)}} \varepsilon_{B,-3}^{\frac{p+2}{2(p+4)}}\varepsilon_{e,-1}^{\frac{2(p-1)}{p+4}}\cr
&&\hspace{1.8cm} \theta_{j,5}^{\frac{2(2-p)}{p+4}}\Delta\theta_{15}^{\frac{4(p-6)-k(p-10)}{p+4}}E_{51}^{\frac{p-2}{p+4}}t_{7.0}^{\frac{16-10k-6p+kp}{2(p+4)}}\cr 
\nu_{\rm a,3}&\simeq& \nu^0_{\rm a,3} \left(\frac{1+z}{1.025}\right)^{\frac{2(4k-5)}{5}}\zeta_{e}^{\frac35}(1+Y)A_{k}^{\frac{8}{5}} \varepsilon_{B,-3}^{\frac{6}{5}}\theta_{j,5}^{-\frac{2}{5}} \Delta\theta_{15}^{\frac{4(4k-5)}{5}}\cr
&&\hspace{5.5cm}\times E_{51}^{\frac{1}{5}} t_{7.0}^{\frac{5-8k}{5}}\,.
\eary
}

The dynamics of the model post the off-axis phase are explored in further detail in \cite{2022arXiv220502459F}.

\section{Polarization model}\label{sec3}

The phenomenon of polarization, the restriction of the vibrations on a wave partially or wholly to a specific geometrical orientation, in GRBs has been observed since 1999 \citep{covino2003polarization}.  Polarization is typically attributed to synchrotron radiation behind the shock waves, which then makes it dependent on the magnetic field configuration and the geometry of the shock, as these will determine the polarization degree ($\Pi$) on each point and its integration over the unresolved image \citep{Gill-1}. The treatment is done by the Stokes parameters, I, Q, U, and V, and typically only linear polarization is considered. From here on forward, we will use the terms unprimed and prime to refer to them in the observer and comoving frames, respectively. In this case,

\begin{align}
    V &= 0, \\
    \frac{U}{I} &= \Pi'\sin{2\theta_p},\\
    \frac{Q}{I} &= \Pi'\cos{2\theta_p},\\
    \theta_p &= \frac{1}{2}\arctan{\frac{U}{Q}}\,,
\end{align}
where $\theta_p$ is the polarization degree. The measured stokes parameters are the sum over the flux \citep{Granot-P2}, so
\begin{align}
    \frac{U}{I} &= \frac{\int \mathrm{d}F_\nu\Pi'\sin{2\theta_p}}{\int \mathrm{d}F_\nu},\\
    \frac{Q}{I} &=  \frac{\int \mathrm{d}F_\nu\Pi'\cos{2\theta_p}}{\int \mathrm{d}F_\nu},
\end{align}
and the polarization is given by
\begin{align}
    \Pi = \frac{\sqrt{Q^2 + U^2}}{I}.
\end{align}

In a thin shell scenario, $\mathrm{d}F_\nu \propto   \delta_D^3L'_{\nu'}\mathrm{d}\Omega$ where $L'_{\nu'}$ is the spectral luminosity and $\mathrm{d}\Omega$ is the element of solid angle of the fluid element in relation to the source.  Using  the approximations $\Tilde{\mu} = \cos{\tilde{\theta}} \approx 1 - \tilde{\theta}^2/2$ and $\beta \approx 1 - 1/2\Gamma^2 $,  $\delta_D$ can be rewritten as $ \delta_D \approx \frac{2\Gamma}{1+\Tilde{\xi}}$, where $\beta$ is the velocity of the material in terms of the speed of light, $\tilde{\theta}$ the polar angle measured from the Line of Sight (LOS) and $\tilde{\xi} \equiv (\Gamma\tilde{\theta})^2$.

Assuming a power-law spectrum and dependency on the $r$, the luminosity can be described as being proportional to the frequency, magnetic field, and direction unity vector \citep{ribicky}
\begin{align}
    L'_{\nu'} \propto (\nu')^{-\alpha} (\sin{\chi'})^\epsilon r^m \propto (\nu')^{-\alpha} (1-\hat{n}' \cdot \hat{B}')^{\epsilon/2} r^m.
\end{align}

We assume, throughout the text, a power-law spectrum and power-law dependency on emissitivy, furthermore we take that the emissivity is radially constant \citep[i.e. $m=0$;][]{Gill-1}. The index $\epsilon$ is dependent on the electron distribution, and we take that $\epsilon=1+\alpha$, where $\alpha$ is the spectral index. The term $\chi$ here is the angle between the local magnetic field and the particle's direction of motion. Since synchrotron emission is highly beamed, the pitch angle is also between the velocity vector and magnetic field. 
The pitch angle, $\chi$, carries the geometric information of the problem, from the structure of the magnetic field ($\hat{B}'$) to the direction of emission ($\hat{n}'$). The geometrical idiosyncrasies of polarization can then be taken in consideration by averaging this factor over the local probability distribution of the magnetic field \citep[see Eq. 15 of][]{Gill-1},

\begin{align}
    \Lambda = \expval{(1-\hat{n}' \cdot \hat{B}')^{\epsilon/2}}.
\end{align}

A Lorentz transformation can be done on either of the unit vectors such as $\hat{n} =\sin{\theta_{\rm obs}}\hat{x} + \cos{\theta_{\rm obs}}\hat{z}$, a normal vector with the direction of the emitting photon in a reference system where the jet axis is in the z-axis,  or a prescription of $\hat{B}$, by using [see, \cite{Lyutikov}]

\begin{align}
    \hat{X}' = \frac{(1+\Gamma)\hat{X}+ \Gamma^2(\hat{X}\cdot\mathbf{v})\mathbf{v}}{(1+\Gamma)\sqrt
    {1+\Gamma^2(\hat{X}\cdot\mathbf{v})}},
\end{align}
so that $\Lambda$ can be expressed in terms of different magnetic field configurations \citep{Gill-1, 2005ApJ...625..263G, Lyutikov, Granot-P2}, as required.

\begin{figure}
{\includegraphics[width=\textwidth]{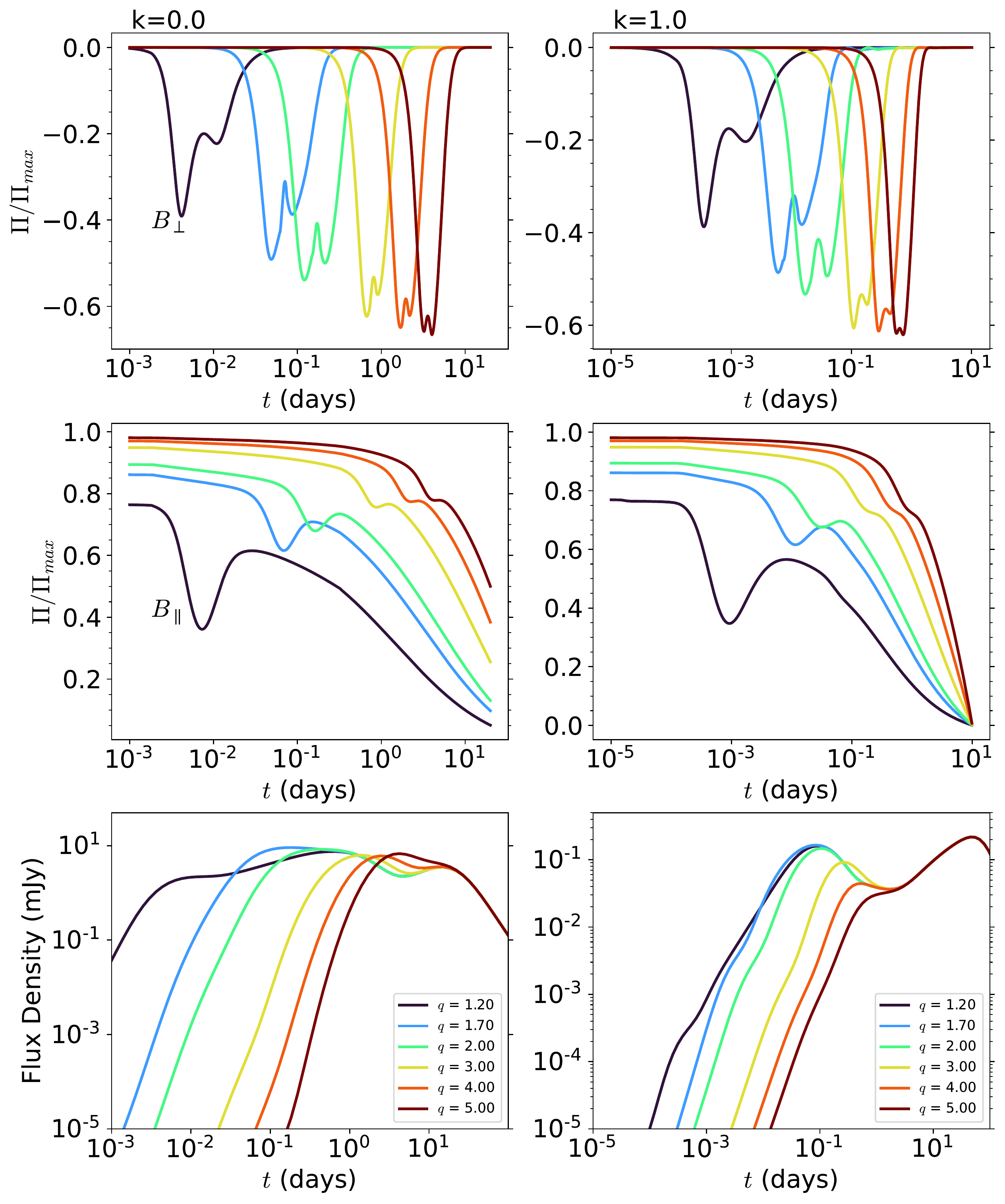}}
\caption{Temporal evolution of the polarization presented for two configurations of magnetic fields - Perpendicular ($B_\perp$) and Parallel ($B_\parallel$), for values of $k=0,\,1$ (from left to right). These polarization curves were calculated for a set of general parameters of GRBs observed in the literature (see \Cref{tab:table_general}). Here, $q_0$ represents the fraction between the observation and the initial opening angle of the jet. The values of $A_k$ are dependent on which values of $k$ are used. These curves may need to be re-scaled by a factor of <1, as {\bf $\Pi_{max}=70\%$ (corresponding to $\alpha=0.60$}) was chosen arbitrarily. To obtain the flux light curves, at the radio frequency of $15.5 \, {\rm GHz}$, the additional parameters of $\varepsilon_B=10^{-4}$, $\varepsilon_e=10^{-1}$, $p=2.2$, $\zeta_e = 1$, initial Lorentz factor $\Gamma_0=100$, and $d_z = 6.6 \, {\rm Gpc} $  were used.  } 
\label{fig:general_case_k0k1}
\end{figure}

{The following equations (see Eqs. 28 and 29 of \cite{Gill-1})

\begin{align}\label{conections}
    \cos{\psi(\tilxi)} = \frac{(1-q)^2 \xi_j - \tilxi}{2q\sqrt{\xi_j\tilxi }} \qquad q = \frac{\theta_{\rm obs}}{\theta_j} \nonumber\\ 
    \xi_j = (\Gamma\theta_j)^2, \qquad \xi_\pm = (1\pm q)^2\xi_j,
\end{align}
regarding the limits of integration of the polarization, can be used to link our synchrotron model to polarization by introducing the bulk Lorentz Factor and the dynamical evolution of the jet's half-opening angle, and thus the physical parameters of the system, obtained in \cite{2022arXiv220502459F}. We want to emphasize that the $q$ parameter evolves with time $q=q(t)=\theta_{\rm obs}/\theta_j(t)$ for a spreading jet.

One of the still-unsolved mysteries of GRBs is the configuration of the magnetic field. As such, various possible configurations must be explored in a topic where magnetic field geometry is of paramount relevance, like polarization. The more used arguments for the symmetry of the magnetic field are varied based on the GRB epoch of relevance for each model. For a scenario where the afterglow is being modeled by a forward shock, two of the most suitable configurations are a random perpendicular configuration, confined to the shock plane (i.e. a field with anisotropy factor {\small $b \equiv \frac{2\expval{B_\parallel^2}}{\expval{B_\perp^2}}= 0$} ) and a parallel configuration along the velocity vector (i.e. a field with anisotropy factor $b\rightarrow\infty$). Here we limit ourselves to these cases ---  an ordered magnetic field parallel to the velocity vector and a random magnetic field generated on the forward shock. However, we would like to add that exploring more complex configurations, such as anisotropic magnetic fields \citep{2020MNRAS.491.5815G, 2021MNRAS.507.5340T, 2020ApJ...892..131S, 2018ApJ...861L..10C} or evolving configurations, is warranted and needed. 

\paragraph{Ordered magnetic field  (parallel configuration)}
The symmetry of the magnetic field configuration causes the polarization to vanish over the image if viewed on-axis ($\theta_{\rm obs} = 0$) or if the beaming cone is wholly contained within the jet aperture. To break the symmetry, the jet must be viewed close to its edge ($q\gtrsim 1+\xi_j^{-1/2}$) where missing emission (from $\theta > \theta_j$) results only in partial cancellation \citep{Waxman}.
For the parallel configuration, the calculation follows Eq. 4 of \cite{Granot-P2}, or using $\Lambda(\tilxi) = \Lambda_\parallel$ from equation 16 of \cite{Gill-1} on Eq. 30 of the same paper.

\paragraph{Random magnetic field  (perpendicular configuration)}
The same symmetry concerns regarding the parallel configuration carry over to the random magnetic field. The equation necessary to calculate this polarization follows equation 5 in \citep{Granot-P2}, or Equation 34 on \cite{Gill-1} when using $\Lambda(\tilxi) = \Lambda_\perp$.

\subsection{Polarization evolution in a stratified medium}
\Cref{fig:general_case_k0k1,fig:general_case_k1.5k2} show the temporal evolution of the polarization degree for the parallel and perpendicular magnetic field configurations and four different possible scenarios to the bulk Lorentz factor  defined with each density profile for $k=0, 1, 1.5 {\rm\,, and\,} 2$. Table \ref{tab:table_general} shows the values utilized to generate these Figures. These generic values are chosen based on the typical ones found for each parameter in the GRB synchrotron literature.  The values of observation angle are varied over a range between $1.2$ and $5$ times the initial opening angle of the jet. This range of values is shown in these figures with different colored lines, each standing for a value of $q_0=\frac{\theta_{\rm obs}}{\theta_{j,0}}$, the ratio between the observation angle and the initial opening angle of the jet. \Cref{fig:joined_q_general} shows the evolution of $q(t)$ for each value of $k$ mentioned above, where $\theta_j(t)$ is associated with the dynamical evolution of the jet \citep[see equations 1 to 4 of][]{2000MNRAS.316..943H}.\footnote{We use a theoretical approach instead of hydrodynamical simulations. See Sec. 2.3 of \cite{2022arXiv220502459F} for the comparison with the hydrodynamical model.} It can be seen that the values of $q$ decline over time and evolves toward $q\rightarrow0$. This is dictated by jet dynamics, as the opening angle of the jet expands as the jet evolves.  By looking at higher values of $q_0$, such as $q_0=4$, we can see from the evolution of this parameter that $q \approx 1.9$ and  $0.8$ at $t=0.9$ and $10$ days, respectively, for $k=0$. The angular evolution of the outflow is essential, as one of the significant issues in polarization is that the fluence drops rapidly for $q>1$, for a top-hat jet where the emission drops sharply past the edges of the jet, which causes difficulties in observing the polarization. This can be easily observed in rows 3 of \Cref{fig:general_case_k0k1,fig:general_case_k1.5k2}, where we present the flux light curves\footnote{The slope variation, circa dozens of days, in the light curve is due to the passage of the synchrotron cooling break through the R-band (15.5 GHz).} at the radio frequency for our chosen parameters. An increase of $q_0$ leads to a decrease of the flux magnitude at earlier times, with the previously mentioned value of $q_0 = 4$ returning an initial flux eight orders of magnitude smaller than the value of $q_0=1.2$, for $k=0$.

\Cref{fig:general_case_k0k1} shows the polarization behavior for the cases with a constant medium -- $k=0$ and  $\Gamma \propto \cos{\Delta\theta}^{\frac{3}{2}}t^{-\frac{3}{2}}$ -- and a stratified medium, with $k=1$ and $\Gamma \propto \cos{\Delta\theta} t^{-1}$. In the perpendicular case, the evolution observed for the $k=0$ scenario presents a distinct polarization peak whose magnitude depends on the geometric parameter $q_0$, a measure of how off-axis the observer is. We can see that the initial polarization for all configurations is at zero. This initial polarization quickly evolves towards a peak once the deceleration timescale ($t_{dec}$) is achieved; the jet expands faster and eventually breaks, which causes the polarization to evolve towards zero after the second peak. Two peaks are present for each value of $q_0$, with the magnitude of the peak increasing with $q_0$ and the peaks merging towards a single peak as $q_0$ increases. This characteristic has been observed before \citep[see][for examples of this dual peak behavior on an off-axis jet polarization case]{2002ApJ...570L..61G, 2004MNRAS.354...86R} and we find our curves to behave similarly.\footnote{The polarization achieved by our model decays faster than for those of the cited works, we believe this is due the chosen evolution of the bulk Lorentz factor and the fact our approach to the evolution of $\theta_j$ has faster increase than the hydrodynamical approach on a timescale of days.} It is believed that each peak is associated with the contribution of the nearest and furthest edges of the jet.
The parallel configuration demonstrates a higher duration on the variability of the polarization -- with a total decreasing behavior across the observation time, but a short interval where a local minimum is generated, with this variance dependent on $q_0$ -- alongside initially high polarization yields.

For the $k>0$ cases, the polarization has been pushed to an earlier time. As such, the evolution starts earlier and peaks earlier. For the sake of clarity, the lower boundary of the x-axis was lowered further. This time behavior happens due to the fact that the afterglow timescale is  {\small $\propto (E/A_k)^{1/(3-k)}$} \citep{Kumar,2003A&A...410..823L, 2022arXiv220502459F}. As such, for the parameters we have chosen for our calculations, the afterglow polarization is shown at earlier times. For more typical parameters, the lower densities of a wind-like medium cause the polarization to evolve slower \citep{2004A&A...422..121L}. The same behavior can be observed in \Cref{fig:general_case_k1.5k2}, where the polarization is presented for the $k=1.5$ -- $\Gamma \propto \cos{\Delta\theta}^{\frac{3}{4}}t^{-\frac{3}{4}}$ -- and $k=2.0$ -- $\Gamma \propto \cos{\Delta\theta}^{\frac{1}{2}}t^{-\frac{1}{2}}$ -- cases.

\section{Polarization from GRBs showing off-axis afterglow emission}\label{sec4}
In this following section, we describe the polarization obtained for a group of GRBs that show similar characteristics on their afterglow: GRB 080503, GRB 140903A, GRB 150101B, GRB 160821B, GRB 170817A \citep[see][for an analysis of the similarities]{fraija-atypical} and SN2020bvc. For a more thorough analysis of the light curves modeling, the Markov Chain Monte Carlo (MCMC) simulations utilized to obtain the parameters used for these calculations, and observation data regarding these bursts, see \cite{2022arXiv220502459F} and the references therein. For this section, we will adopt the notation $f(q_0=x^{\pm y}_{\pm z}) = a^{\pm b}_{\pm c}$.

\paragraph{GRB 080503}
The first column in \Cref{fig:joined_GRB} shows the theoretical polarization evolution calculated for GRB 080503 for the magnetic field configurations --- perpendicular and parallel, from top to bottom, respectively.  The parameters for calculating this polarization are presented on the first row of \Cref{tab:pol_grbs}. A negligible influence of the value of $q_0$ is observed on the peak polarization for both configurations, with peak polarization $\abs{\Pi}(B_\perp) \approx 41\%$ and initial polarization $\Pi(B_\parallel) \approx 65\%$. For the perpendicular field, the somewhat small effect of $q_0$ can be observed on the peak time, with $t_{peak}(q_0=2.37^{+0.05}_{-0.05}) \approx 2.15^{+0.28}_{-0.21}$ days and null polarization is reached at $\sim 15$ days --- the second peak is observed at $t \approx 3.8^{+0.5}_{-0.4}$ with $\abs{\Pi} \approx 30\%$. The local minimum polarization of the parallel magnetic field configuration is observed at  $t \approx 2.71^{+0.34}_{-0.31}$ days, with a magnitude of $\Pi(q_0=2.37^{+0.05}_{-0.05}) \approx 52^{+0.2}_{-0.4}\%$ and a $\sim 2\%$ increase is observed after $\sim 1.6^{+0.2}_{-0.1}$ days. After that event, the polarization decreases steadily to $\Pi (t=100) \approx 13.5^{+0.5}_{-0.5}\%$.

\paragraph{GRB 140903A}
The second column in \Cref{fig:joined_GRB} shows the theoretical polarization evolution estimated for GRB 140903A, similarly to the previous case. The parameters are presented on the second row of \Cref{tab:pol_grbs}. The chosen value of $q_0$ shows a higher degree of influence for this burst, even if changed just slightly.  The perpendicular case shows a peak polarization of  $\abs{\Pi}(q_0=1.61^{+0.08}{-0.08}) \approx 33^{+1.2}_{-1.0}\%$
at the times $t_{peak}(q_0=1.61^{+0.08}{-0.08}) \approx 4.1^{+1.7}_{-1.2}\times10^{-2}$ days. The second peak manifests at $t= 6.5^{+1.7}_{-1.9}\times10^{-2}$ days, with magnitudes of $\abs{\Pi} \approx 24.9^{+0.7}_{-1.0}\%$, and zero polarization is reached at $\approx 0.4$ days. For the parallel configuration the initial polarization is $\Pi \approx 58.2^{+1.0}_{-1.2}\%$. The local minimum is $\Pi \approx 41.7^{+1.4}_{-2.0}\%$ at the times $t= 5.8^{+2.4}_{-1.7}\times10^{-2}$ days and a increase of $\approx 8^{+1}_{-1}\%$ is observed after $7^{+3}_{-1}\times10^{-2}$ days before steady decline. A polarization of $\Pi \approx 1\%$ is observed at the 100 day mark.

\paragraph{GRB 150101B}
The third column in \Cref{fig:joined_GRB} shows the theoretical polarization calculated for GRB 150101B. The parameters for calculating this polarization are presented on the third row of \Cref{tab:pol_grbs}. In a similar manner to GRB 080503, the different values of $q_0$ offer at best a differential change on the polarization. For the perpendicular case we observe the following: peak polarization of $\abs{\Pi} \approx 38\%$ at $t_{peak}(q_0=2.08^{+0.04}_{-0.04}) \approx 4.8^{+0.2}_{-0.5}$ days, with zero reached at $13$ days --- the second peak is observed at $t \approx 3.8^{+0.5}_{-0.4}$ with $\abs{\Pi} \approx 28\%$. For the parallel case we see that the initial polarization is $\abs{\Pi} \approx 63\%$, the local minimum is observed at  $t = 2.15^{+0.25}_{-0.20}$ days, with a magnitude of $\Pi(q_0=2.08^{+0.04}_{-0.04}) \approx 48.8^{+0.2}_{-0.5}\%$, and a $\sim 4\%$ increase is observed after $\sim 2.2^{+0.2}_{-0.5}$ days. After that event, the polarization decreases steadily to $\Pi (t=100) \approx 16^{+1}_{-1}\%$ .

\paragraph{GRB 160821B}
The fourth column in \Cref{fig:joined_GRB} shows the theoretical evolution of polarization calculated for GRB 160821B. The parameters for calculating this polarization are presented on the fourth row of \Cref{tab:pol_grbs}. 
These polarization curves behave more similarly to the ones observed in GRB 140903A, with some peculiarities. 
The perpendicular case shows a peak polarization of  $\abs{\Pi}(q_0=1.27^{+0.13}{-0.13}) \approx 28.4^{+1.6}_{-2.9}\%$
at the times $t_{peak}(q_0=1.27^{+0.08}_{-0.08}) \approx 2.9^{+5.1}_{-2.4}\times10^{-2}$ days. The second peak is fairly prominent, showing at  $t = 5.2^{+5.4}_{-4.2}\times10^{-2}$ days, with magnitude of $\abs{\Pi} \approx 21^{+2}_{-6}\%$. The polarization eventually reaches zero at $t = 1.6^{+3.7}_{-1.2}\times10^{-1}$ days.
For the parallel case, a pulsation of small magnitude ($\lesssim 1\%$) is observed at the initial period of time, where the polarization is expected to decrease softly with our fiducial model, for $q_0=(1.27, 1.40)$. This pulsation is not observed for $q_0=1.14$ likely due to the fact that a smaller value of $q_0$ pushes the polarization faster in time, causing it to happen before our lower time boundary. Overall, the polarization at initial times is $\Pi\approx 54.8^{+2.0}_{-2.2}\%\pm 1\%$. The local minimum polarization is $\Pi \approx 30^{+6}_{-8}\%$ at $t_{peak} = 4.6^{+8.6}_{-3.7}\times 10^{-2}$ days, a 
 $\sim 17^{-5}_{+7}\%$ increase is observed after $\sim 8.6^{+11.7}_{-6.6}\times 10^-2$ days. After which, the polarization steadily decreases to $\Pi(t=100) \approx 3^{+1}_{-1}\%$.

\paragraph{GRB 170817A}
\Cref{fig:joined_170817A_b} shows the expected polarization, calculated with our model, for the different configurations of magnetic fields. An extensive analysis of the synchrotron light curves was done by \cite{2019ApJ...884...71F},  where the authors have fitted the synchrotron light curves with a dual component model, and we aim to expand this analysis to the polarization. The off-axis component dominates the late afterglow period after two weeks \citep[see][]{2017Sci...358.1559K, 2017MNRAS.472.4953L, mooley, 2018ApJ...867...95H, 2019ApJ...884...71F}; thus we only use the off-axis component, an expanding top-hat jet, to calculate the polarization. A similar approach was done by \cite{2021MNRAS.507.5340T}, whom also used a dual component outflow --- albeit with a structured jet.  We have used the values reported in Table 1 of \cite{2019ApJ...884...71F} to generate the polarization curves.  The synchrotron analysis done for GRB 170817A was calculated for the scenario where $k=0$, and the same condition is applied to our model. As such, the polarization presents a similar behavior as the left side of \Cref{fig:general_case_k0k1}.

First, we see across the different configurations that the chosen array of observation angles, chosen based on MCMC simulations, leads to a granular increment of $q_0$ that has little to no effect on the overall polarization evolution. As such, we will limit ourselves to the analysis of a single value of $q_0=3.28$. For the perpendicular case, we see that the polarization is initially null and shows a rapid increase to a peak of $\abs{\Pi} \approx 46\%$ at $t\approx 141$ days and declines to zero again when $t\approx 432$ days, where it remains. The parallel case has an initially high polarization of $\Pi \approx 68\%$ that decreases softly until a sharper decrease happens at $t\approx 180$ days and the polarization becomes $\Pi \approx 54\%$. A small increase of $\sim 1\%$ happens again at $t \approx 240$ days from where the polarization starts to decrease sharply, reaching $\sim 33\%$ at $t \approx 10^3$ days}. \cite{2018ApJ...861L..10C} report an upper limit of $\Pi < 12\%$, with $99\%$ confidence at $2.8$ GHz and  $t_{\rm obs} \approx 244$ days. Our results for both configurations of magnetic fields return values  that infringe on the upper limits. As such, based on our model of jet dynamics, presented in further detail in \cite{2022arXiv220502459F} and \cite{2019ApJ...884...71F}, we can rule out the fully anisotropic scenario.

We want to highlight that some authors \cite[e.g., see][]{2018MNRAS.478.4128G, 2020ApJ...892..131S, 2020MNRAS.491.5815G, 2021MNRAS.507.5340T}  have already tried to constrain the magnetic field configuration using the polarization upper limit, from radio observations, for this particular burst.   \cite{2018MNRAS.478.4128G} have calculated the polarization for a gaussian jet, power-law jet, and quasi-spherical outflow with energy injections for three anisotropy values, $b=(0.0,\, 0.5,\,1.5)$.  They have found that the structured jets produce a high polarization degree ($\Pi\approx60\%$, for $b=0$) peaking at $\sim 300$ days, with the wide-angle quasi-spherical outflow with energy injection returning a lower polarization degree ($\Pi\approx10\%$) at all times, with all values of $b$.  \cite{2021MNRAS.507.5340T} and \cite{2020MNRAS.491.5815G} have obtained the polarization for different anisotropy factors and found, for the dynamical evolution dictated by their jet models, that a random magnetic field should be close to isotropic (b=1) to satisfy the polarization upper limits. \cite{2021MNRAS.507.5340T} also expanded that a magnetic field with two components, an ordered and a random, could satisfy the upper limits should $0.85<b<1.16$ and the ordered component be as high as half the random one. \cite{2020ApJ...892..131S} have analyzed the measured and theoretical polarization ratio for a non-spreading top-hat off-axis jet that constrains the geometry of the magnetic fields to a dominant perpendicular component, but with a sub-dominant parallel component ($b>0$). 
Of these authors, \cite{2021MNRAS.507.5340T, 2020MNRAS.491.5815G} and \cite{2018MNRAS.478.4128G} have explored the $b=0$ scenario for GRB 170817A. They reported  $\Pi\approx60\%$ at $\sim 300$ days, with the polarization still decreasing softly for times upwards of $10^3$ days. While our polarization values and evolution are somewhat different, likely due to different synchrotron models and parameters, it remains that our explored cases have also broken the available upper limits, ruling out the $b=0 (b\rightarrow\infty)$ possibilities.

More observations on a shorter post-burst period would be needed to  constrain the magnetic field configuration further. Unfortunately, there were no polarization observations at any other frequency and time  \citep{2018ApJ...861L..10C}.

\begin{figure}
{\includegraphics[width=\textwidth]{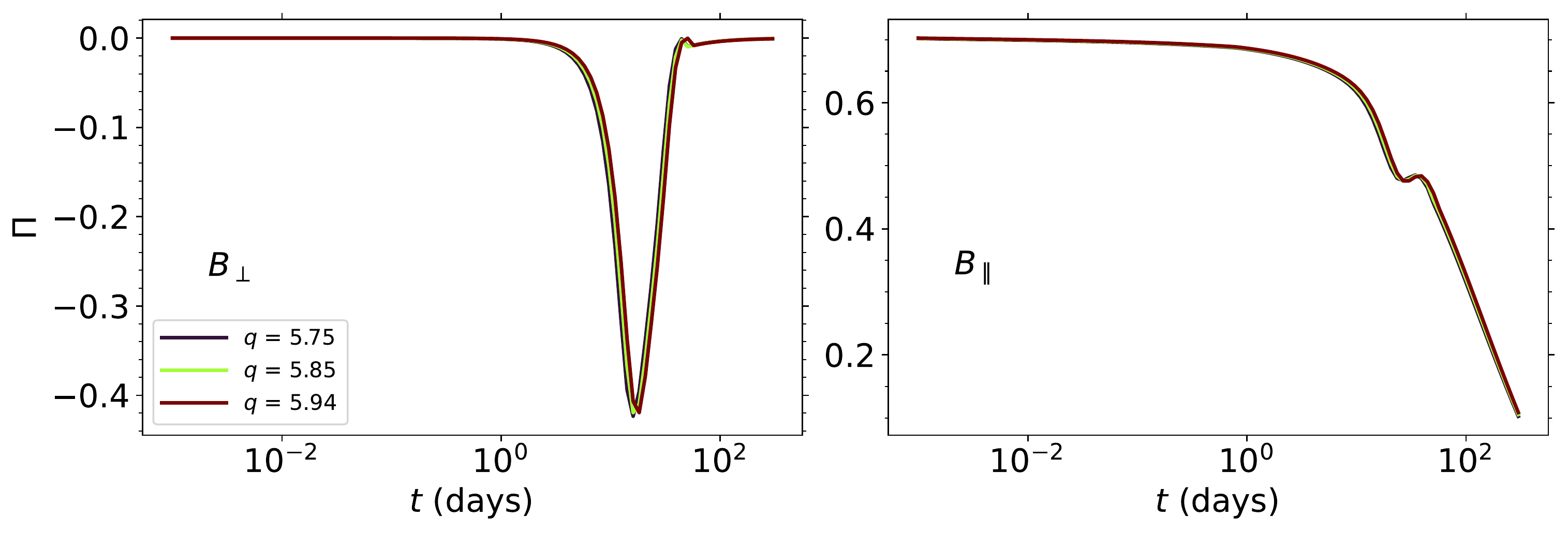}}
\caption{Expected temporal evolution of the polarization for the supernova SN2020bvc, calculated for 3 possible magnetic field configurations - Perpendicular ($B_\perp$), Parallel ($B_\parallel$) and Toroidal ($B_{tor}$), respectively, from left to right. The set of parameters presented on \Cref{tab:par_mcmc} were used, with the uncertainty of the observation angle, $\theta_{\rm obs}$, used to return a range of values for the fraction $q_0$, represented by the colormap legend. For this case, $k=1.5$.}
\label{fig:joined_supernova}
\end{figure}

\paragraph{SN 2020bvc}
\Cref{fig:joined_supernova} shows the expected polarization for SN 2020bvc calculated for a stratified medium where $k=1.5$. The parameters used to calculate the values of polarization are presented on the fifth row of \Cref{tab:pol_grbs}. The perpendicular case shows a peak polarization of  $\abs{\Pi}(q_0=5.85^{+0.09}_{-0.10}) \approx 42^{-1}_{+1}\%$
at the times $t_{peak}(q_0=1.61^{+0.08}_{-0.08}) \approx 16.1^{+2.0}_{-0.1}$ days, with a null polarization state at $44$ days. The parallel case, on the other hand, has a initial maximum polarization of  $\Pi \approx 70\%$, a local minimum of $\Pi \approx 47 \%$ at $t_{peak} = 26$ days, a $\sim 1\%$ increase is observed after $\sim 8$ days. After which, the polarization steadily decreases to $\Pi(t=100) \approx 10\%$.

\section{Conclusions}\label{sec7}

We have introduced a polarization model as an extension of the analytical synchrotron afterglow off-axis scenario presented in \cite{2019ApJ...884...71F, 2022arXiv220502459F}.  We have shown this model's expected temporal polarization evolution, dependent on the physical parameters associated with afterglow GRB emission. This synchrotron model describes the multiwavelength afterglow for homogeneous and stratified ambient media based on the parameter $k$ ($k=0$ for homogeneous and $k>0$ for stratified). The polarization allows us to speculate on the nature of the magnetic field, which originates the synchrotron flux on the afterglow. We have calculated the polarization for a broad set of parameters, constrained within the typical values observed for off-axis GRBs, for four different stratification states  ($k=[0, 1, 1.5, {\rm\, and\,} 2]$) and the two magnetic field configurations. We assumed a wholly perpendicular configuration contained to the shock plane (i.e., the anisotropy factor $b=0$) or a wholly ordered configuration parallel to the shock normal (i.e., the anisotropy factor $b\rightarrow \infty$) .

For these simulations, we were able to distinctly see the difference in possible polarization caused by the stratification of the ambient medium for both field configurations. The perpendicular magnetic field configuration shows prominent peaks whose magnitude becomes increasingly higher as the observer is further away from the edge of the jet. The parallel configuration, on the other hand, showed initially high polarization yields with a local minimum observed, before a regrowth and eventual decrease towards zero as the jet laterally expands. The influence of the $q_0$ ratio is evident, as the initial polarization is higher with an increasing $q_0$, but the magnitude of the local minimum decreases inversely with $q_0$. This influence of the observation angle on the peak of the polarization is a result that agrees with the polarization literature \citep{1999MNRAS.309L...7G, Granot-P2, 2004MNRAS.354...86R, Gill-1}. The effect of stratification on the polarization seems to be two-fold, one result coming from typical GRB behavior, where the afterglow timescale is proportional to the inverse of the density -- $t\propto A_k^{-1/(3-k)}$, as such higher or lower densities push the polarization to different timescales; the second result comes at the magnitude of a discontinuity observed at the time of the jet-break, with this ``polarization break"  becoming increasingly higher with the stratification parameter.

We have also obtained the expected polarization curves for a sample of bursts showing off-axis afterglow emission - GRB 080503, GRB 140903A, GRB 150101B, GRB 160821B, GRB 170817A, and SN2020bvc.  In particular, we have used the available polarimetric upper limits of GRB 170817A, $\abs{\Pi} < 12\%$ at 2.8 GHz and $t\approx244$ days \citep{2018ApJ...861L..10C}, in a attempt to constrain the magnetic field geometry. The polarization obtained with our jet dynamics and the chosen anisotropy returns a value that  breaks the established upper limits on both of the configurations, which in turn allow us to rule out the $b=0(b\rightarrow \infty)$ cases. 

\begin{figure}
{\includegraphics[width=\textwidth]{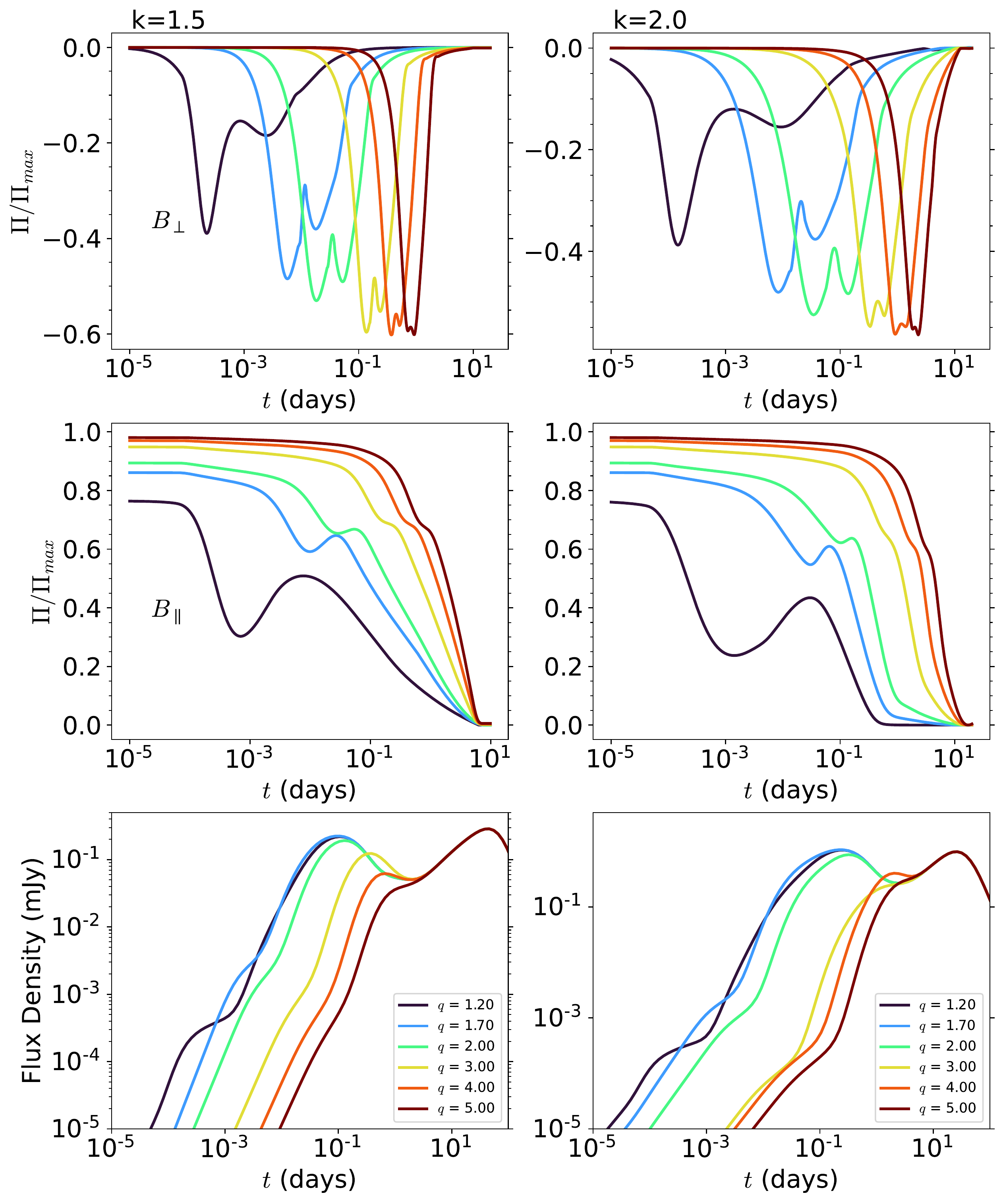}}
\caption{Temporal evolution of the polarization presented for two configurations of magnetic fields - Perpendicular ($B_\perp$) and Parallel ($B_\parallel$), for values of $k=1.5,\,2$ (from left to right). These polarization curves were calculated for a set of general parameters of GRBs observed in the literature (see \Cref{tab:table_general}). Here, $q_0$ represents the fraction between the observation and the initial opening angle of the jet. The values of $A_k$ are dependent on which values of $k$ are used. These curves may need to be re-scaled by a factor of <1, as $\Pi_{max}=70\%$ (corresponding to $\alpha=0.60$) was chosen arbitrarily. To obtain the flux light curves, at the radio frequency of $15.5 \, {\rm GHz}$, the additional parameters of $\varepsilon_B=10^{-4}$, $\varepsilon_e=10^{-1}$, $p=2.2$, $\zeta_e = 1$, initial Lorentz factor $\Gamma_0=100$, and $d_z = 6.6 \, {\rm Gpc} $ were used.}
\label{fig:general_case_k1.5k2}
\end{figure}

\begin{figure}
{\includegraphics[width=\textwidth]{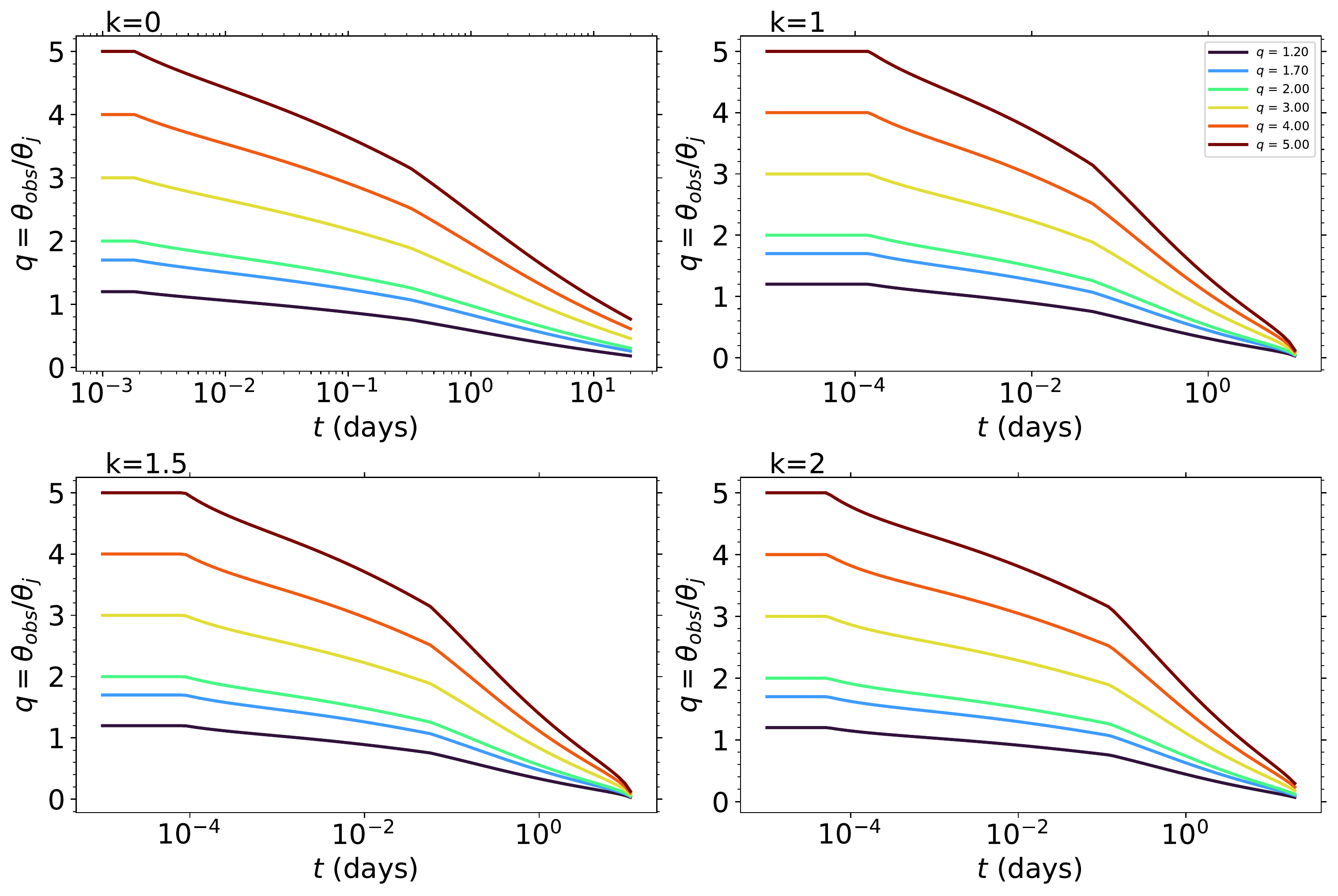}}
\caption{Temporal evolution of the $q$ parameter, for the four stratification scenarios presented on \Cref{fig:general_case_k0k1,fig:general_case_k1.5k2}.}
\label{fig:joined_q_general}
\end{figure}

Although the remaining bursts have neither detected polarization nor constrained upper limits to compare with, analysis of these bursts that appear to show similar nature can be of use in the occasion more similar bursts are found in the future. From our calculations we have observed the following similarities:

For the perpendicular field configuration, GRB 080503 and GRB 150101B show somewhat similar magnitudes of polarization at similar times. GRB 140903A and GRB 160821B also present some similarities on their polarization magnitudes, but here a higher difference on the time at which the peaks are displayed is present, with the lowest value of $q_0$ used for GRB 160821B having a polarization peak one order of magnitude earlier in time. In all likelihood this differentiation between the two groups of bursts comes from the angular properties of the jet, as for the latter group the initial value of $q_0$ is closer to unity. Furthermore, the differences between GRB 140903A and GRB 160821B likely also come from angular properties, as they become more amplified for even small changes in $q_0$, as $q_0$ is close to unity.  The polarization obtained, with our model, for GRB 170817A is closer to that presented for the former group than the latter. With the peaks showing as $\abs{\Pi} \approx 41\%$ at $\sim 2.15$ days (GRB 080503), $\abs{\Pi} \approx 38\%$ at $\sim 4.8$ days (GRB 150101B) and $\abs{\Pi} \approx 46\%$ at $\sim 141$ days (GRB 170817A).
SN2020bvc is modeled in a stratified medium, $k=1.5$, unlike the bursts mentioned above. As such, the expected polarization should be similar to the left side of \Cref{fig:general_case_k1.5k2} and that holds true. However, a particularity of our modelling of SN2020bvc is that the initial value of $q_0$ is incredibly high, which in turn  leaves the polarization in a similar state to the $k=0$ scenario for similarly high values of $q_0$. For all these bursts the peak of polarization has roughly coincided with the peak of the flux for an off-axis observer \citep[see][for the flux fitting]{2022arXiv220502459F}, which is a result that agrees with the literature \citep{1999MNRAS.309L...7G, 2003ApJ...594L..83G, 2004MNRAS.354...86R, 2021MNRAS.507.5340T}.
Overall, we can see that certain similarities can be observed between the bursts' polarizations. However, the peculiarities of each burst make so none are the same.
More observations on durations from seconds to months after the trigger are needed to infer tighter constraints on polarization and proper fitting of the flux data needed to dissolve the degeneracy between models.

\begin{acknowledgments}
We thank Walas Oliveira,  Rodolfo Barniol Duran, Tanmoy Laskar, Paz Beniamini and Bing Zhang for useful discussions. The authors would also like to extend their gratitude towards the peer-reviewer in charge of this manuscript, for his invaluable input and contributions to the betterment of this work. AP acknowledges financial support from CONACyT's doctorate fellowships, NF acknowledges financial  support  from UNAM-DGAPA-PAPIIT  through  grant IN106521. RLB acknowledges support from CONACyT postdoctoral fellowships and the support from the DGAPA/UNAM IG100820 and IN105921.
\end{acknowledgments}

\bibliographystyle{aasjournal}
\bibliography{references}  
\newpage

\clearpage
\begin{table}
\centering \renewcommand{\arraystretch}{1.85}\addtolength{\tabcolsep}{1pt}
\caption{Constants of the relevant quantities in the synchrotron scenario from Section \ref{subsec21}}
\label{Table1}
\begin{tabular}{l   c  c  c c }
 \hline \hline
 &  ${\bf k=0}$    &\hspace{0.5cm}   ${\bf k=1.0}$  &\hspace{0.5cm}   ${\bf k=1.5}$  &\hspace{0.5cm}   ${\bf k=2.0}$     \\ 
\hline
$A_{\rm k}$ & $1\,{\rm cm^{-3}}$ & $ 10^{17}\,{\rm cm^{-2}}$ & $3.1\times 10^{25}\,{\rm cm^{-\frac32}}$ & $10^{34}\,{\rm cm^{-1}}$ \\
\hline
%
%
$\gamma^0_{\rm m}\,  (\times 10)$ & $1.15\times10^{-1}$ &	$1.36\times10^{-3}$ &	$5.16\times10^{-4}$ &	$2.85\times10^{-4}$  \\
$\gamma^0_{\rm c}\, (\times 10^{4})$ & $1.53\times 10^{1}$ &	$7.46\times 10^{-3}$ &	$1.59\times 10^{-3}$ &	$6.20\times 10^{-4}$  \\
$\nu^{\rm 0}_{\rm a,1}\,(\times 10^{-6}\, \rm Hz)$ & $9.73\times 10^{-4}$ &	$1.32\times 10^{+2}$ &	$1.50\times 10^{+3}$ &	$6.58\times 10^{+3}$  \\
$\nu^0_{\rm a,2}\,(\rm \times10^{-5}\,Hz)$ & $2.96\times 10^{-2}$ &	$2.46\times 10^{-1}$ &	$3.82\times 10^{-1}$ &	$4.98\times 10^{-1}$  \\
$\nu^0_{\rm a,3}\,(\rm \times10^{-5}\, Hz)$ & $1.29\times 10^{1}$ &	$7.21\times 10^{4}$ &	$4.61\times 10^{5}$ &	$1.43\times 10^{6}$  \\
$\nu^{0}_{\rm m}\,(\times10^{-6}\,\rm Hz)$ & $2.52\times 10^{-4}$ &	$1.08\times 10^{-6}$ &	$3.24\times 10^{-7}$ &	$1.54\times 10^{-7}$  \\
$\nu^0_{\rm c}\,(\times10^{-1}\,\rm Hz)$ & $4.42\times 10^{1}$ &	$3.25\times 10^{-4}$ &	$3.06\times 10^{-5}$ &	$7.31\times 10^{-6}$  \\
$F^0_{\rm max}\,(\times10\,{\rm mJy}) $ & $1.27$ &	$3.90\times 10^{1}$ &	$8.12\times 10^{1}$ &	$1.27\times 10^{2}$  \\
\hline
\end{tabular}
\end{table}


\begin{table}
\centering \renewcommand{\arraystretch}{2.2}\addtolength{\tabcolsep}{2.4pt}
    \caption{Table of values used to obtain the Polarization curves for the general case}
    \label{tab:table_general}
    \begin{tabular}{l c c c c} 
     \hline
     \hline
     & $\mathbf{k=0}$ & $\mathbf{k=1.0}$ & $\mathbf{k=1.5}$ & $\mathbf{k=2.0}$\\
     \hline
     $A_k$ & $1\,{\rm cm}^{-3}$ & $ 10^{19}\,{\rm cm}^{-2}$ & $2.6 \times10^{27}\, {\rm cm}^{-3/2}$ & $ 10^{36}\,{\rm cm}^{-1}$ \\
      $E$($10^{52}$ erg) & 5 & 5 & 5 & 5\\
      $\theta_j$(deg) & $4$ & $4$ & $4$ & $4$\\
      $\theta_{\rm obs}$(deg) & $[1.2,5]\theta_j$ & $[1.2,5]\theta_j$ & $[1.2,5]\theta_j$ & $[1.2,5]\theta_j$ \\
      \hline
      \hline
    \end{tabular}
    \\
    \centering{The range [$1.2,5$] for $\theta_{\rm obs}$ represents the interval $[1.2, 1.7, 2.0, 3.0, 4.0, 5.0$]}
    \\
\end{table}

\begin{table}
\centering \renewcommand{\arraystretch}{2.2}\addtolength{\tabcolsep}{2.4pt}
\caption{Median Values of Parameters used to calculate the polarization curves for a sample of short and long GRBs.}\label{tab:par_mcmc}
\begin{tabular}{ l c c c c c c}
\hline
\hline\\ 
{\large   Parameters} & {\large $E\, (10^{52}\,{\rm erg})$} & {\large ${\rm A_0}\,\, (10^{-2}\,{\rm cm^{-3}})$} & {\large $A_{1.5}\,\, (10^{22}\,{\rm cm^{-3/2}})$} & {\large $\theta_{\rm j}\,\,(\rm deg)$} & {\large $\theta_{\rm obs}\,\,(\rm deg)$} & {\large p}\footnote{The electron power-law index, p, returns the spectral index by taking $\alpha = \frac{p-1}{2}$} \\
\hline \hline
\\
{\large GRB 080503} & $2.156^{+0.294}_{-0.295}$ & $4.221^{+0.102}_{-0.103}$ & -- & $6.589^{+0.081}_{-0.078}$ & $15.412^{+0.268}_{-0.269}$ & $2.319^{+0.049}_{-0.049}$  \\
{\large GRB 140903A} & $3.163^{+0.290}_{-0.296}$ & $4.219^{+0.102}_{-0.101}$ & -- & $3.210^{+0.080}_{-0.081}$ & $5.162^{+0.271}_{-0.267}$ &  $2.073^{+0.048}_{-0.050}$ \\
{\large GRB 150101B} & $1.046^{+0.120}_{-0.124}$ & $0.164^{+0.021}_{-0.021}$ & -- & $6.887^{+0.662}_{-0.682}$ & $14.114^{+2.327}_{-2.179}$ &  $2.150^{+0.217}_{-0.215}$  \\
{\large GRB 160821B} & $0.118^{+0.021}_{-0.021}$ & $0.869^{+0.093}_{-0.090}$ & -- & $8.002^{+0.817}_{-0.809}$ & $10.299^{+1.125}_{-1.135}$ &  $2.220^{+0.021}_{-0.021}$ \\
{\large SN 2020bvc} & $0.238^{+0.011}_{-0.010}$ & -- & $9.984^{+0.195}_{-0.193}$ & $2.121^{+0.078}_{-0.079}$ & $12.498^{+0.268}_{-0.281}$  & $2.313^{+0.037}_{-0.035}$ \\
\hline
\label{tab:pol_grbs}
\end{tabular}
\end{table}


\clearpage

\clearpage

\begin{figure}
{\includegraphics[width=\textwidth]{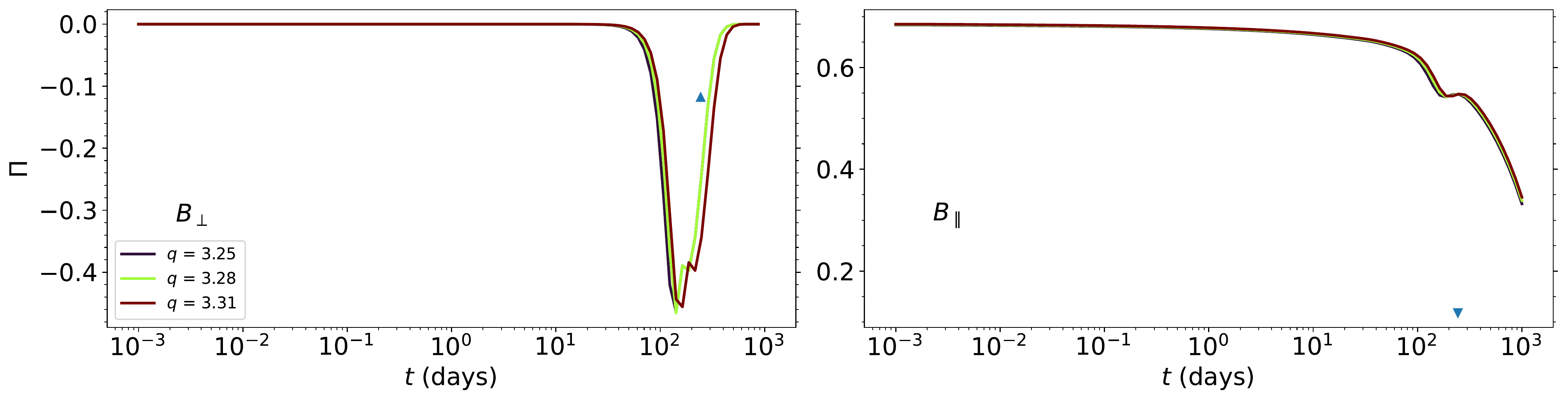}}
\caption{Expected Temporal evolution of the polarization for GRB 170817A for two configurations of magnetic fields - Perpendicular ($B_\perp$), Parallel ($B_\parallel$). These polarization curves were calculated using the best fit values presented in Table 1 of \cite{2019ApJ...884...71F}:
$\tilde{E} \approx 6.3 \times 10^{49}$erg, $n \approx 2.8 \times 10^{-4} {\rm cm}^{-3}$, $\theta_{\rm obs} \approx [24.5, 24.7, 25.0]$ and $\theta_j=7.6$ deg, $p \approx 2.24$. The blue triangles represent the Radio upper limit of $\abs{\Pi}=12\%$. Radio upper limit at 2.8 GHz was taken from \cite{2018ApJ...861L..10C}.} 
\label{fig:joined_170817A_b}
\end{figure}
\clearpage




\begin{figure}
{\includegraphics[width=\textwidth]{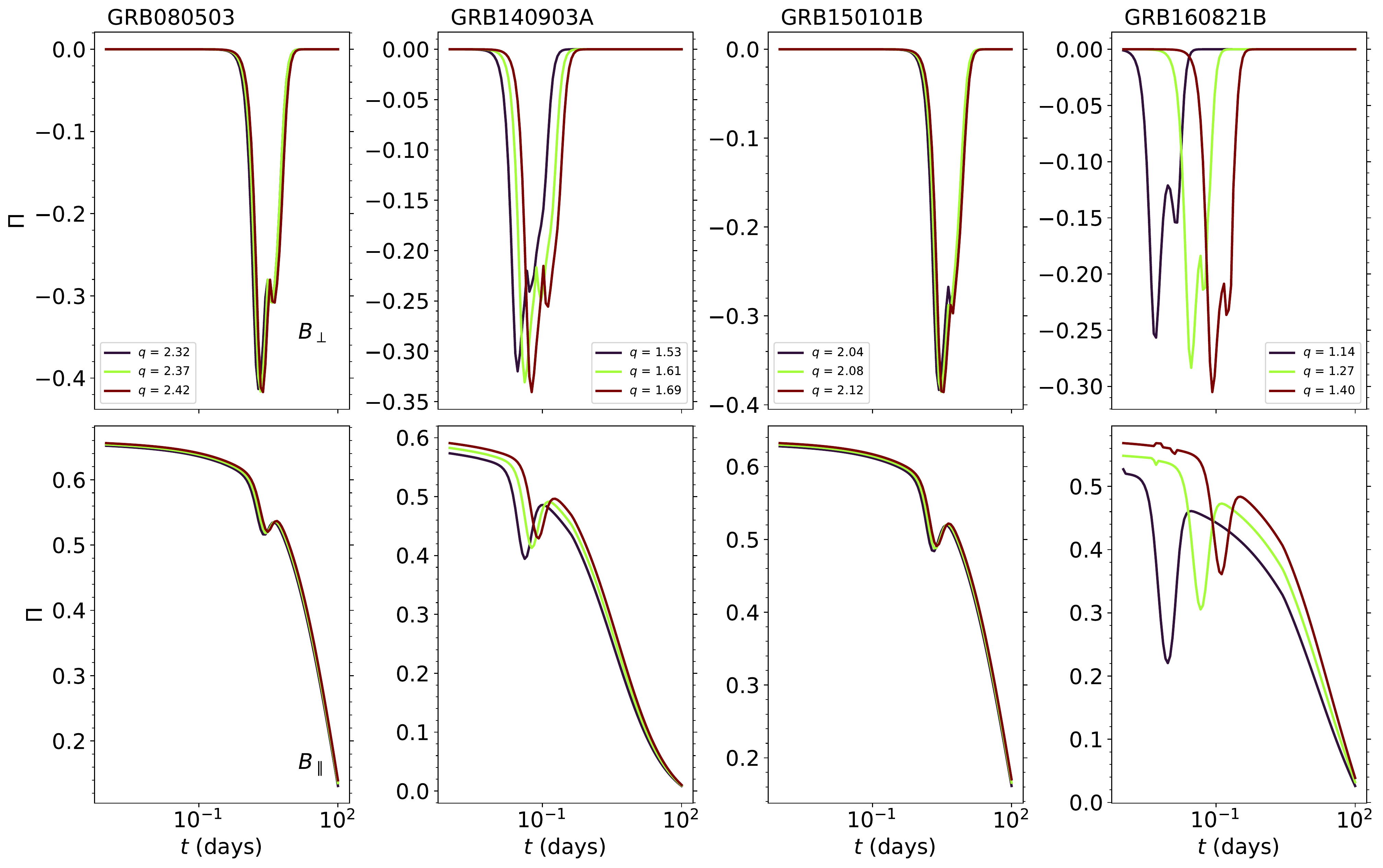}}
\caption{Expected temporal evolution of the polarization for the bursts GRB 080503, GRB140903, GRB150101B, GRB160821B - respectively, from left to right. Each burst has its polarization calculated for 3 possible magnetic field configurations - Perpendicular ($B_\perp$), Parallel ($B_\parallel$) and Toroidal ($B_{tor}$), respectively from top to bottom - for the set of parameters presented on \Cref{tab:par_mcmc}. The uncertainty of the observation angle, $\theta_{\rm obs}$, was used to return a range of values for the fraction $q_0$, represented by the colormap legend on the figure. All polarizations are calculated for the case $k=0$.}
\label{fig:joined_GRB}
\end{figure}


\clearpage

\end{document}